\documentclass[twocolumn]{aastex62}

\usepackage{graphicx}

\newcommand{\micro}{$\mathrm{\mu}$}
\newcommand{\los}{$_{\mathrm{los}}$~}

\submitjournal{ApJ}
\accepted{July 25, 2019}

\shorttitle{Survey of Magnetic Field Strengths}
\shortauthors{Thompson et al.}

\begin{document}

\title{A Survey of Magnetic Field Strengths in the Envelopes of Molecular Clouds via the 18 cm OH Zeeman Effect }

\correspondingauthor{K.L. Thompson}
\email{krthompson@davidson.edu}

\author{K.L. Thompson}
\affil{Davidson College, 209 Ridge Road, Davidson, NC 28035, USA}

\author{T.H. Troland}
\affil{University of Kentucky, Lexington, KY 40506}

\author{C. Heiles}
\affil{University of California, Berkeley, CA 94720}

\begin{abstract}
We present the results of an extensive Arecibo observational survey of magnetic field strengths in the inter-core regions of molecular clouds to determine their role in the evolution and collapse of molecular clouds as a whole. Sensitive 18~cm~OH Zeeman observations of absorption lines from Galactic molecular gas in the direction of extragalactic continuum sources yielded 38 independent measurements of magnetic field strengths.  Zeeman detections were achieved at the three sigma level toward 9 clouds, while the others revealed sensitive upper limits to the magnetic field strength.  Our results suggest that total field strengths in the inter-core regions of GMCs are about 15 \micro{}G.

\end{abstract}

\keywords{ISM: magnetic fields - stars: formation}

\section{Introduction} \label{sec:intro} 

It has long been known that stars form in the gravitational collapse of an interstellar molecular cloud, but the details of the process are not yet well understood \citep{Cru12}. The evolution of self-gravitating molecular clouds depends upon the ratio of the internal energies of support to external energies of confinement.  In the absence of internal support mechanisms, a molecular cloud would undergo gravitational collapse and form stars on the free-fall timescale.  This would lead to a Galactic star formation rate of approximately 250~M$ _{\odot}$~yr$^{-1}$, which is far greater than the observed formation rate of~$\sim$~3~M$ _{\odot}$~yr$^{-1}$ \citep{McK99}.  Therefore, molecular clouds appear to be forming stars inefficiently, suggesting that there must be some means of internal support that is hindering the gravitational collapse of molecular clouds and lengthening cloud lifetimes beyond the free-fall timescale.  Any successful theory of star formation must, therefore, account for this inefficiency.  Two prevailing theories of star formation have emerged, one placing emphasis on the support provided by magnetic fields, and the other on turbulence.

The magnetically driven model of star formation suggests that ambipolar diffusion plays a crucial role in the evolution of molecular clouds \citep{Mou99, Shu99, McK99}.  In this model, molecular clouds that are supported by magnetic fields at formation will, over time, become unstable and collapse to form low-mass stars through the ambipolar diffusion process.  

The turbulence driven model suggests that star formation is driven by supersonic turbulence within the clouds \citep{Mac04}.  In this theory, clouds form intermittently at the intersections of supersonic flows in the interstellar medium (ISM).  Usually, these clouds dissipate, but occasionally they can become gravitationally bound and collapse to form stars.  This theory accounts for the inefficiency of star formation in that only a small fraction of the clumps become self-gravitating and undergo collapse.  Magnetic fields are present in this theory, but they are weak and do not affect the overall process of star formation.  

To distinguish between these two theories of star formation, it is necessary to determine the mass-to-flux ratio, M/$\Phi$, within molecular clouds to determine if these clouds are gravitationally dominated (supercritical) or magnetically dominated (subcritical).  The ambipolar diffusion model of star formation predicts that M/$\Phi$ should be subcritical in the envelopes of molecular clouds and slightly supercritical in their cores.  If magnetic fields are too weak to be important, as the turbulence model implies, then one would expect M/$\Phi$ to be supercritical.  

There have been many studies aimed at determining the mass-to-flux ratio in molecular clouds.  \citet{Cru99} summarized the available results of measurements of M/$\Phi$ in molecular cores and HI regions near newly formed stars, and concluded that M/$\Phi$ is supercritical by a factor of approximately 2.  Subsequent studies of the mass-to-flux ratio in cloud cores by \citet{Bou01}, \citet{Tro08}, and \citet{Fal08} revealed no clear examples of subcritical molecular clouds and found that cloud cores are slightly supercritical, in agreement with \citet{Cru99}.  An extensive survey of HI Zeeman splitting measurements was undertaken by \citet[and references therein]{Hei05}, resulting in magnetic field strengths and column densities for diffuse regions of the ISM in the direction of 79 continuum sources.  Mass-to-flux ratios derived from this ``Millennium Survey'' suggest that the diffuse material probed by HI lies in the subcritical regime. 

Previous Zeeman effect studies of molecular clouds apply mainly to cloud cores.  However, the cores represent only a small fraction (of order 10\%) of GMC masses \citep{Bat14, Hey15}.  Here, we present Zeeman effect measurements of 18~cm OH absorption lines along random lines-of-sight through GMCs to determine the magnetic flux within the inter-core regions of the clouds.  These lines-of-sight are defined by the locations of background extragalactic continuum sources.  This project consumed about 400 hours of Arecibo Observatory telescope time.  From these measurements, we can begin to understand the role of magnetic fields in GMCs as a whole, rather than in the cores alone.   In \S 2 and 3 we discuss target selection and observations.  In \S 4, we review the Zeeman effect method for direct detection of magnetic field strengths.  We present the results and their analysis in \S 5, and summarize these results and discuss future work in \S 6.

\section{Selection of Targets} \label{sec:targets}

Potential targets were selected from extra-galactic continuum sources that lie behind galactic molecular clouds within the Arecibo declination range.  Molecular clouds were identified from the CO maps of \citet{Dam01}, and bright (S$_{\nu} >$~0.5~Jy) continuum sources spatially coincident with suitable molecular clouds were selected using the 1.4~GHz NRAO VLA Sky Survey (NVSS) \citep{Con98}.  A high level of sensitivity is required to detect the Zeeman effect in radio frequency spectral lines.  Therefore, we chose extragalactic background sources toward which there are strong, narrow CO emission lines in the Dame et al. survey.  We were, of course, limited by the declination range of the Arecibo telescope (0 - 40 degrees), and we excluded sources for which OH lines had already been observed in the Millennium Survey.  Final targets were selected from among those having strong and narrow OH absorption lines as observed at Arecibo.  Sensitivity requirements are such that integrations of several 10s of hours (see \textbf{Table \ref{tab:sources}}) were often needed to achieve $\sigma(B_{los})$ of 5~$\mu$G, where B\los is the line-of-sight field strength.

In addition to targets observed as part of this study, we include 5 targets observed in OH as part of the Millenium Survey \citep[and references therein]{Hei05} which are found to lie behind molecular gas in the CO maps of \citet{Dam01} but for which magnetic field strengths were not previously determined. Our final sample consists of 21 lines-of-sight through molecular clouds.  In general, absorption spectra show many individual velocity components due to the existence of multiple molecular clouds along a single line-of-sight, each allowing for an independent calculation of field strength and column density.  This is especially true for low-latitude sources in the direction of the Galactic center.  As a result, we yield 38 independent Zeeman magnetic field measurements from our 21 line-of-sight observations.  
  
The targets for this study are diverse, in that they are distributed between the Galactic center (R.A.~$\approx$~19$^h$~-~21$^h$) and the Galactic anti-center (R.A.~$\approx$~03$^h$~-~07$^h$) regions of the sky, allowing us to probe variations in physical properties between the two regions.  In addition, most targets lie in the direction of nearby molecular clouds associated with low-mass star formation (e.g.,~Taurus), but a few allow us to sample clouds known for high-mass star formation (e.g.,~Monocerous OB1).  The lines-of-sight through molecular clouds probed by this sample are not biased toward molecular cores, as were previous observations, since extra-galactic continuum sources are distributed at random throughout the sky and the probability that the line-of-sight of our observations pass through molecular cores is small.

\section{Observations} \label{sec:obs}
% check the 400 hour thing - is this accurate?
The Arecibo telescope was used between 2009 September and 2012 June to conduct Zeeman observations of the four 18~cm ground states of OH at 1612, 1665, 1667, and 1720~MHz.  To maximize the efficiency of the telescope and achieve the highest Zeeman sensitivity, in-band frequency switching was used for the majority of the observations.  However, a small amount of time was allotted to measure OH emission lines off-source.  These data, along with the on-source absorption profiles, can be used to derive OH excitation temperatures and column densities \citep{Hei03a, Li18}.

Simultaneous Zeeman observations of the four OH 18~cm lines were carried out using the L-band wide receiver with native linear polarizations, and correlation methods were used to derive the Stokes parameters.  This correlation technique is described by \citet{Hei04}.  The correlator sampled 2048 channels in each frequency band over a spectral bandwidth of either 0.78 or 1.56~MHz, depending upon the absorbing velocity range of individual target sources, resulting in OH main line velocity resolutions of 0.068~km~s$^{-1}$ or 0.137~km~s$^{-1}$ for the 0.78 and 1.56~MHz bandwidths, respectively.  \textbf{Table \ref{tab:sources}} provides the name, location, flux as seen by the NVSS, on-source integration time for the OH main lines, and the OH bandwidth used for each target source.  

In addition to our target sources for Zeeman observations, we briefly observed the well-known maser region W49(OH).  Stokes I and V spectra toward W49(OH) were compared to those of \citet{Col70} to verify the sense of circular polarization, and thus the magnetic field direction. We also observed S88B, a well-known and well-studied galactic HII region associated with active star formation with a strong Zeeman effect to verify Zeeman field calculations.

\begin{deluxetable}{cccccc}
\tabletypesize{\footnotesize}
\tablecaption{Target Sources}
\tablewidth{0pt}
\tablehead{\colhead{Source}	& \colhead{$\mathrm{\ell}$} 	& \colhead{\textit{b}} 		& \colhead{S$_{\nu}$\tablenotemark{a}} 	& \colhead{t$_{int}$}	& \colhead{Bandwidth}  \\
\colhead{} & \colhead{} & \colhead{} & \colhead{(Jy)} & \colhead{(hr)} & \colhead{(MHz)} }
\startdata
3C092						& 159.7		& -18.4		& 1.6			& 23.2		& 0.78			\\
3C123\tablenotemark{b}		& 170.6		& -11.7		& 49.7			& 4.1		& 1.56			\\
3C131						& 171.4		& -7.8		& 2.87			& 27.0		& 0.78			\\
3C133\tablenotemark{b}		& 177.7		& -9.9		& 5.8			& 5.0		& 0.78			\\
3C154\tablenotemark{b}		& 185.6		& 4.0		& 5.0			& 11.1		& 0.78			\\
3C207\tablenotemark{b}		& 213.0		& 30.1		& 2.6			& 27.5		& 0.78			\\
3C417				& 73.3		& -5.5		& 4.8			& 14.8		& 0.78			\\
4C+13.67			& 43.5		& 9.2		& 1.6			& 19.1		& 0.78			\\
4C+14.18			& 197.0		& 1.1		& 2.4			& 24.9		& 0.78			\\
4C+17.23			& 176.4		& -24.2		& 1.0			& 16.6		& 0.78			\\
4C+27.14			& 175.8		& -9.4		& 0.9			& 19.1		& 0.78			\\
B0531+2730			& 179.9		& -2.8		& 1.0			& 19.2		& 0.78			\\
B1853+0749			& 40.5		& 2.5		& 3.3			& 16.6		& 0.78			\\
B1858+0407			& 37.8		& -0.2		& 2.1			& 14.4		& 1.56			\\
B190840+09			& 43.3		& -0.8		& 8.7			& 13.9		& 1.56			\\
B1919+1357			& 48.9		& -0.28		& 4.5			& 15.6		& 1.56			\\
B1920+1410 			& 49.2		& -0.34		& 6.9			& 14.8		& 1.56			\\
B2008+3313			& 71.2		& -0.09		& 1.6			& 15.8		& 0.78			\\
PKS0528+134			& 191.4		& -11.0		& 1.6			& 21.5		& 0.78			\\
S88B				& 61.5		& 0.1		& 4.0			& 9.9		& 0.78			\\
T0629+10\tablenotemark{b}	& 201.5		& 0.5		& 2.4			& 7.2		& 0.78			\\
\enddata       
\tablenotetext{a}{Source fluxes taken from the NVSS \citep{Con98}}    
\tablenotetext{b}{Observed as part of the Millennium Survey \citep{Hei03a,Hei03b,Hei04,Hei05}} 
\label{tab:sources}
\end{deluxetable}

\section{Zeeman Effect} \label{sec:zeeman}

The radio frequency Zeeman effect provides the only known method to measure magnetic field strengths directly in localized regions of the ISM.  The effect amounts to a frequency offset between a spectral line observed in opposite senses of circular polarization.  In practice, the frequency offset is a small fraction of the line width, and it is detected in the Stokes V profile as a scaled-down replica of the derivative of the Stokes I profile.  In this limit of small frequency offset, the amplitude in Stokes V of the scaled-down Stokes I derivative is proportional to $\mathrm{B_{los}}$.  Therefore, B\los can be derived by fitting the Stokes V profile to the derivative of the Stokes I profile (or to the analytical derivative of a Gaussian function fitted to the Stokes I profile, an approach we adopt here).  The fitting process yields $\mathrm{B_{los}}$, $\sigma(\mathrm{B_{los}})$, and the sign of $\mathrm{B_{los}}$, with positive values indicating a field directed away from the observer.  This fitting process is described by, among other authors, \citet{Cru93} and \citet{Sar13}.  Note that line profiles with multiple velocity components can yield multiple independent measures of $\mathrm{B_{los}}$.  Also, the two OH main lines (1665 and 1667 MHz) yield independent measures of $\mathrm{B_{los}}$.  

The OH satellite lines (1612 and 1720 MHz) exhibit a more complex Zeeman effect.  Also, these lines are usually very non-thermal in excitation, leading to complex line profiles.  Therefore, the OH satellite lines are generally not useful for Zeeman effect measurements of the type described here.

\section{Analysis and Results} \label{sec:analysis}

Magnetic field strengths were determined independently for the 1665~and~1667~MHz OH lines, and a single field was computed by taking the weighted average of the two measurements, weighted by the inverse square of the 1$\sigma$ uncertainly in the field.  Results of our Zeeman analysis can be found in \textbf{Table~\ref{tab:Bresults}} with 1$\sigma$ uncertainties.  The magnetic field strength determined from the 1665~MHz line is shown in column 3, the field strength from the 1667~MHz line in column 4, and the weighted mean line-of-sight field for the two OH main lines in column 5.  

The observed Stokes I absorption profiles in the direction of several target sources contained multiple components which were not well-separated in velocity space.  Velocities for these individual line components are given in column 2.  Only Gaussian components for which the uncertainty in B$_{\mathrm{los}}$, $\sigma_{B}$, was less than 20~\micro{}G and the 1665~and 1667~MHz measurements agree within error ($|$B$_{1665}$~-~B$_{1667}|$~$<$~$\sigma_{B}$) were included in the table and used for subsequent analysis.  

We find that the magnetic field satisfies the condition $|$B$_{\mathrm{los}}|$~$>$~3$\sigma_{B}$ in the direction of 9 velocity components belonging to the lines-of-sight toward background sources 3C092, 3C123, 3C133, 3C154, 4C+13.67, B1853+0479, B1858+0407, and B1919+1357.  Field strengths for these components are shown in boldface in \textbf{Table~\ref{tab:Bresults}}.

\begin{deluxetable*}{ccccc}[h!]
	\tabletypesize{\footnotesize}
	\tablecaption{Magnetic Field Results}
	\tablewidth{0pt}
	\tablehead{\colhead{Source}	& \colhead{V$\mathrm{_{lsr}}$} 	& \colhead{B$\mathrm{_{los,65}}$} 		& \colhead{B$\mathrm{_{los,67}}$} 	& \colhead{B$\mathrm{_{los,ave}}$}	 \\
		\colhead{} & \colhead{(km s$^{-1}$)} & \colhead{(\micro G)} & \colhead{(\micro G)} & \colhead{(\micro G)}  }
	\startdata
	3C092		& 8.75 $\pm$ 0.01		& -18.9 $\pm$ 3.7	& -10.4 $\pm$ 4.5	& \textbf{-15.5 $\pm$ 2.8}	   	\\
	3C123		& 5.47 $\pm$ 0.01		& -11.8 $\pm$ 7.9	& -4.6 $\pm$ 5.5	& -7.0 $\pm$ 4.5				\\
				& 4.45 $\pm$ 0.01		& -10.0 $\pm$ 5.3	& -4.4 $\pm$ 4.0	& -6.5 $\pm$ 3.2				\\
				& 3.71 $\pm$ 0.03		& -23.8 $\pm$ 14.7	& -35.4 $\pm$ 13.7	& \textbf{-30.0 $\pm$ 10.0}		\\
	3C131		& 7.24 $\pm$ 0.01		& 2.2 $\pm$ 3.0		& -0.6 $\pm$ 3.4	& 1.0 $\pm$ 2.2					\\
				& 6.59 $\pm$ 0.01		& 10.6 $\pm$ 7.1	& 6.5 $\pm$ 8.2		& 8.9 $\pm$ 5.4					\\
				& 4.61 $\pm$ 0.03		& -10.9 $\pm$ 21.2	& -29.0 $\pm$ 20.7	& -20.1 $\pm$ 14.8				\\
	3C133		& 7.69 $\pm$ 0.01		& -7.9 $\pm$ 2.8	& -4.4 $\pm$ 2.4	& \textbf{-5.9 $\pm$ 1.8}		\\
	3C154		& -2.34 $\pm$ 0.01		& 22.7 $\pm$ 10.3	& 20.2 $\pm$ 10.2	& \textbf{21.5 $\pm$ 7.2}				\\
	3C207		& 4.55 $\pm$ 0.01		& -3.1 $\pm$ 12.3	& -22.4 $\pm$ 12.4	& -12.7 $\pm$ 8.8				\\
	3C417		& 9.80 $\pm$ 0.01		& 4.2 $\pm$ 3.5		& 5.6 $\pm$ 3.4		& 4.9 $\pm$ 2.4					\\
	4C+13.67	& 5.43 $\pm$ 0.03		& 8.6 $\pm$ 11.0	& 11.6 $\pm$ 20.8	& 9.3 $\pm$ 9.7					\\	
				& 4.67 $\pm$ 0.04		& 11.8 $\pm$ 4.8	& 16.4 $\pm$ 9.4	& \textbf{12.7 $\pm$ 4.2}		\\
	4C+14.18	& 32.29 $\pm$ 0.08		& 4.6 $\pm$ 6.0		& 1.8 $\pm$ 6.1		& 3.2 $\pm$ 4.3					\\
	4C+17.23	& 11.23 $\pm$ 0.02		& -10.1 $\pm$ 12.5	& -10.7 $\pm$ 12.4	& -10.4 $\pm$ 8.8				\\
				& 9.16 $\pm$ 0.02		& 3.1 $\pm$ 13.1	& 8.0 $\pm$ 11.3	& 5.9 $\pm$ 8.6					\\
	%4C+27.14	& 7.83 $\pm$ 0.03		& 17.8 $\pm$ 25.5	& -1.3 $\pm$ 10.0	& 1.3 $\pm$ 9.3					\\
	B0531+2730	& 2.91 $\pm$ 0.01		& 7.7 $\pm$ 5.2		& 9.0 $\pm$ 5.1		& 8.3 $\pm$ 5.1					\\
	B1853+0749	& 28.16 $\pm$ 0.03		& 13.8 $\pm$ 2.8	& 10.3 $\pm$ 3.1	& \textbf{12.2 $\pm$ 2.1}		\\
				& 26.66 $\pm$ 0.05	 	& 21.8 $\pm$ 6.9	& 14.3 $\pm$ 3.5	& \textbf{15.8 $\pm$ 3.1}		\\
				& 8.09 $\pm$ 0.01		& 13.1 $\pm$ 5.5	& 5.0 $\pm$ 3.7		& 7.5 $\pm$ 3.0					\\
	B1858+0407	& 20.49 $\pm$ 0.03		& -2.9 $\pm$ 3.3	& -12.5 $\pm$ 4.8	& -6.0 $\pm$ 2.7				\\
				& 19.20 $\pm$ 0.18		& -4.0 $\pm$ 13.3	& -5.5 $\pm$ 6.1	& -5.3 $\pm$ 5.6				\\
				& 16.83 $\pm$ 0.11		& 10.0 $\pm$ 9.7	& -13.5 $\pm$ 10.9	& -0.4 $\pm$ 7.2				\\
				& 14.98 $\pm$ 0.19		& 0.9 $\pm$ 8.1		& -8.4 $\pm$ 6.7	& -4.6 $\pm$ 5.2				\\
				& 13.58 $\pm$ 0.08		& 8.8 $\pm$ 3.5		& 8.8 $\pm$ 5.8		& \textbf{8.8 $\pm$ 3.0}		\\
	B190840+09	& 43.22 $\pm$ 0.03		& 4.5 $\pm$ 9.6		& 10.0 $\pm$ 15.0	& 6.1 $\pm$ 8.1				\\
				& 40.58 $\pm$ 0.03		& 3.6 $\pm$ 4.7		& 4.3 $\pm$ 10.1	& 3.7 $\pm$ 4.3					\\
	B1919+1357	& 6.27 $\pm$ 0.01		& 7.3 $\pm$ 2.8		& 4.2 $\pm$ 2.3		& \textbf{5.4 $\pm$ 1.8}				\\
	B1920+1410 	& 6.31 $\pm$ 0.01		& 4.1 $\pm$ 4.4		& 4.3 $\pm$ 3.2		& 4.2 $\pm$ 2.6					\\
				& 5.11 $\pm$ 0.01		& 3.7 $\pm$ 5.4		& 6.5 $\pm$ 4.7		& 5.3 $\pm$ 3.5					\\
	B2008+3313	& 11.31 $\pm$ 0.02		& 0.79 $\pm$ 3.6	& 4.0 $\pm$ 3.4		& 2.5 $\pm$ 2.5					\\
				& 9.16 $\pm$ 0.03		& 5.0 $\pm$ 6.7		& 9.1 $\pm$ 6.2		& 7.2 $\pm$ 4.6					\\		
	PKS0528+134	& 9.58 $\pm$ 0.02		& -5.9 $\pm$ 7.1	& -1.2 $\pm$ 6.8	& -3.4 $\pm$ 4.9				\\
	T0629+10	& 6.94 $\pm$ 0.03		& 20.1 $\pm$ 9.5	& 16.8 $\pm$ 10.5	& 18.6 $\pm$ 7.0				\\
				& 6.10 $\pm$ 0.02		& 9.2 $\pm$ 6.0		& 3.6 $\pm$ 7.6		& 7.0 $\pm$ 4.7					\\
				& 4.63 $\pm$ 0.01		& 7.1 $\pm$ 4.7		& 4.3 $\pm$ 4.9		& 5.8 $\pm$ 3.4					\\
				& 2.19 $\pm$ 0.01		& -4.8 $\pm$ 2.6	& -3.1 $\pm$ 3.6	& -4.2 $\pm$ 2.1				\\
	\enddata  
	\tablecomments{Magnetic field strengths shown bold satisfy the criterion $|$B$_{\mathrm{los}}|$~$>$~3$\sigma_{B}$.} 
	\label{tab:Bresults}    
\end{deluxetable*}

\begin{figure}
	\figurenum{1}
	\plotone{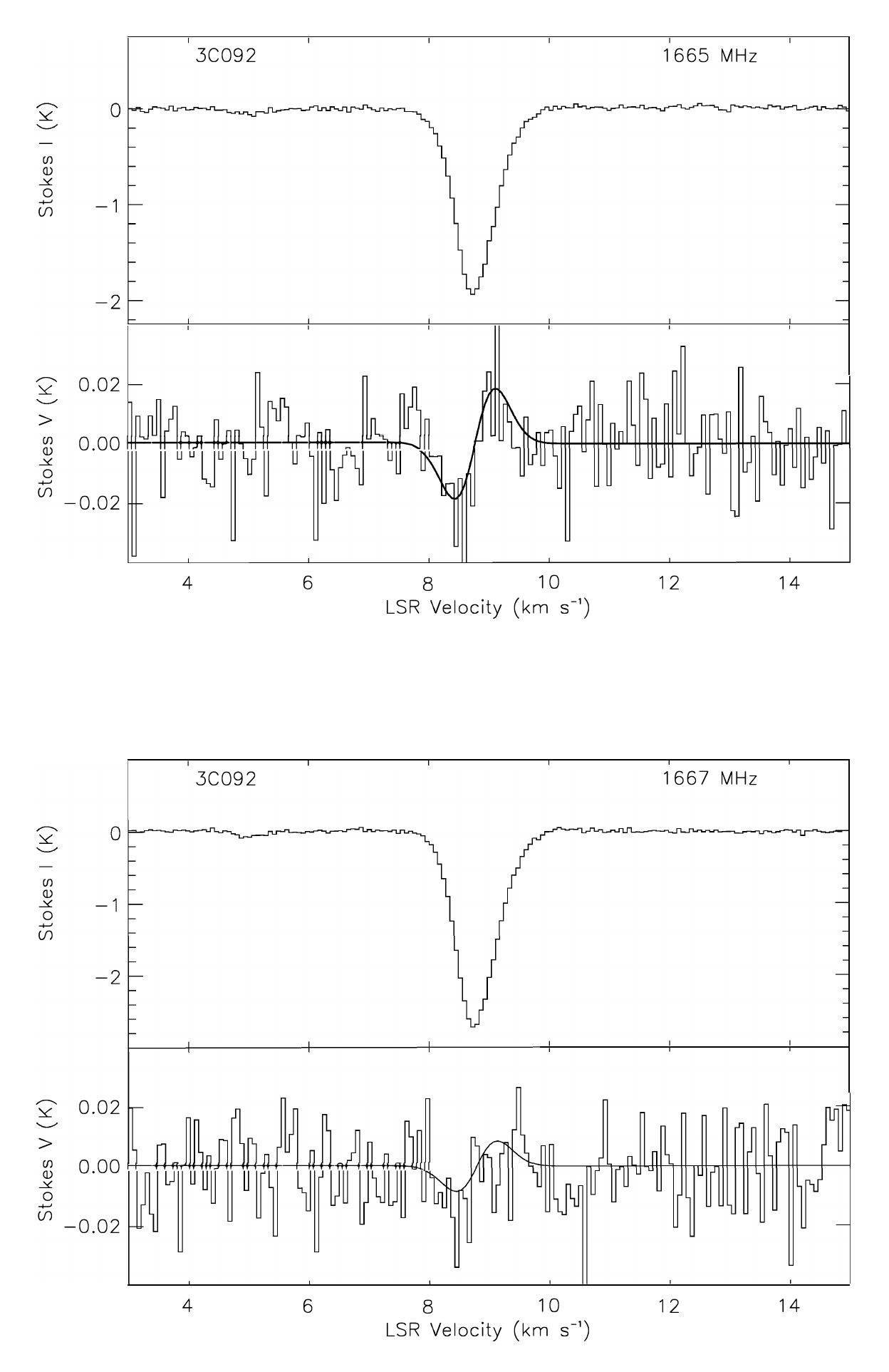}
	\caption{The 1665 MHz (top) and 1667 MHz (bottom) OH spectra.  Stokes I is shown in the top panel and Stokes V in the bottom.  In each figure, the data are shown as histograms and the smooth line in the Stokes V spectrum is the fit. \label{fig:3C092}}
\end{figure}

An example of the Stokes I and V profiles fit for the Zeeman effect is shown in \textbf{Figure~\ref{fig:3C092}} for 3C092, a source with a clear Zeeman detection of B\los~=~-15.5~$\pm$~2.8~\micro{}G.  The top panel shows the Stokes I profile, which was fit with a single Gaussian, and the bottom panel shows the Stokes V profile (histogram) and the least-squares fit of dI/d$\nu$ scaled to reveal the field strength (smooth line).  Similar plots for all observed lines-of-sight included in this study are available in the electronic edition of {\it The Astrophysical Journal}.

Our results for magnetic field strengths are consistent with, and improve upon, previous results from other authors for the same lines-of-sight.   S88B is a source with a well-known Zeeman effect used to verify our Zeeman calculations.  We find a clear Zeeman signal with B$_{los}$~=~53.1~$\pm$~1.4~$\mu$G, which is in close agreement with values of 47~$\pm$~3~$\mu$G and 49~$\pm$~2~$\mu$G reported by \citet{Goo89} and \citet{Cru00}, respectively.  

\citet{Cru81} carried out 1665 and 1667~MHz~OH Zeeman observations toward 3C123 and 3C133.  For 3C123, they report field magnitudes of 11.8~$\pm$~9.7, 6.5~$\pm$~9.2, and 16.8~$\pm$~31.0~\micro{}G for the 5.5, 4.5, and 3.7~km~s$^{-1}$ components, respectively.  We find fields of $|$B$_{\mathrm{los}}|$~=~7.0~$\pm$~4.5, 6.5~$\pm$~3.2, and 30.0~$\pm$~10.0~\micro{}G for the same components.  For 3C133, the authors find a mean field of $|$B$_{\mathrm{los}}|$~=~13~$\pm$~10~\micro{}G, and we report $|$B$_{\mathrm{los}}|$~=~5.9~$\pm$~1.8~\micro{}G.

It is important to note that since the Zeeman effect only reveals the line-of-sight magnetic field strength B$_{\mathrm{los}}$, our determinations of B are lower limits to the total field strength, and a statistical analysis of the results is necessary.  We calculate the mean value of B\los weighted by the inverse square of the uncertainties $\sigma_B$ such that more sensitive measurements of the field receive higher weight in the average.  The mean quantity is derived under the assumption that all magnetic field strengths are the same, and the observed variation in B\los is due only to differing orientations of the field vector for each of our sampled velocity components.  In determining the mean field, we exclude the known HII region S88B since it is not representative of the envelopes of molecular clouds.  However, we include Cassiopeia A (Cas A), for which field strengths and OH column densities were determined by \citet{Hei86}.  The line-of-sight toward this source intersects molecular clouds in the Perseus arm of the Milky Way. We include all magnetic field results in Table~\ref{tab:Bresults} in our mean value, regardless of the uncertainty, since all Zeeman results are useful in the statistical calculation of the weighted mean quantity.  Overall, we find the mean field to be  $\langle{\mathrm{B}}_{\mathrm{los}}\rangle$~=~7.4~$\pm$~0.4~\micro{}G.  We also consider separately the mean field strengths in the Galactic center and Galactic anti-center regions of the sky.  We find that the average field in each region individually is consistent with the overall mean value and equal to 7.0~$\pm$~0.7 and 7.6~$\pm$~0.5~\micro{}G, respectively.  

It is possible to apply a statistical correction to the measured values of B\los to determine a statistically valid total field strength.  For a large ensemble of Zeeman measurements for which the field is randomly oriented, the total field strength is expected to be twice the average B\los \citep{Cru99}.  Therefore, our results suggest that B$_\mathrm{tot}$ is of order 15~\micro{}G in regions of GMCs that lie outside the cores.  If this value is representative of magnetic fields in the inter-core regions of molecular clouds, then it can be compared with B$_{\mathrm{tot}}$~=~6~\micro{}G derived by \citet{Hei05} for the Cold Neutral Medium (CNM).  The CNM is thought to be the precursor to molecular clouds.  Therefore, the scaling relationship of field strength to gas density between the two phases of ISM may offer clues to the evolutionary process.  Heiles~\&~Troland take a mean n(H) of 54~cm$^{-3}$ for the CNM.  Also, \citet{Rom10} find a median n(H) of 230~cm$^{-3}$ for molecular clouds in the Galactic Ring.  Simple application of a scaling law with B proportional to n(H)$^{\kappa}$ implies $\kappa$ about equal to 0.6.  This value, of course, must be viewed with caution since its statistical significance is uncertain.

\section{Summary} \label{sec:sum}

We have used the Arecibo telescope to conduct observations of OH absorption in the direction of 21 extragalactic continuum sources that lie behind Galactic molecular clouds.  Magnetic field strengths within the inter-core regions of these clouds, where few Zeeman studies have focused, were determined via the Zeeman effect.  We detect a line-of-sight magnetic field B\los above the 3$\sigma$ level in 9 velocity components, shown in boldface in Table~\ref{tab:Bresults}.  However, due to the statistical nature of Zeeman effect observations, all field measurements are meaningful in a determination of a weighted mean field value.  We find the mean line-of-sight and total magnetic field strengths to be 7.4~\micro{}G and 14.8~\micro{}G, respectively, in the sampled envelopes of molecular clouds.

The importance of the magnetic field in the support and evolution of GMCs as a whole is not currently understood. One way to reveal the true state of magnetic support within molecular clouds is to determine the ratio of mass to the magnetic flux (M/$\Phi$) within molecular clouds.  The determination of field strengths presented here is the first step in the process.  Further work is required to combine these field results with estimates of OH column density to obtain mass-to-flux ratios for the inter-core regions of molecular clouds.  The mass-to-flux ratio results will be presented in a forthcoming paper.

\acknowledgements

We thank Arecibo for the generous amount of observing time for the completion of this project and the observatory staff for the successful completion of our observations.  We also thank Richard Crutcher for his help in the analysis of our data.  

This work was supported in part by the National Science Foundation under NSF grant AST 0908841. 

The Arecibo Observatory is operated by SRI International under a cooperative agreement with the National Science Foundation (AST-1100968), and in alliance with Ana G. M\'endez-Universidad Metropolitana, and the Universities Space Research Association.

\facility{Arecibo}

\appendix

\section{Astrophysical Journal online-only figures}

The 1665 MHz (top) and 1667 MHz (bottom) OH spectra, to be published only electronically in the ApJ as an extension to Figure 1, are shown here.  Stokes I is shown in the top panel and Stokes V in the bottom.  In each figure, the data are shown as histograms and the smooth line in the Stokes V spectrum is the fit.

\begin{figure}[!h]
	\plotone{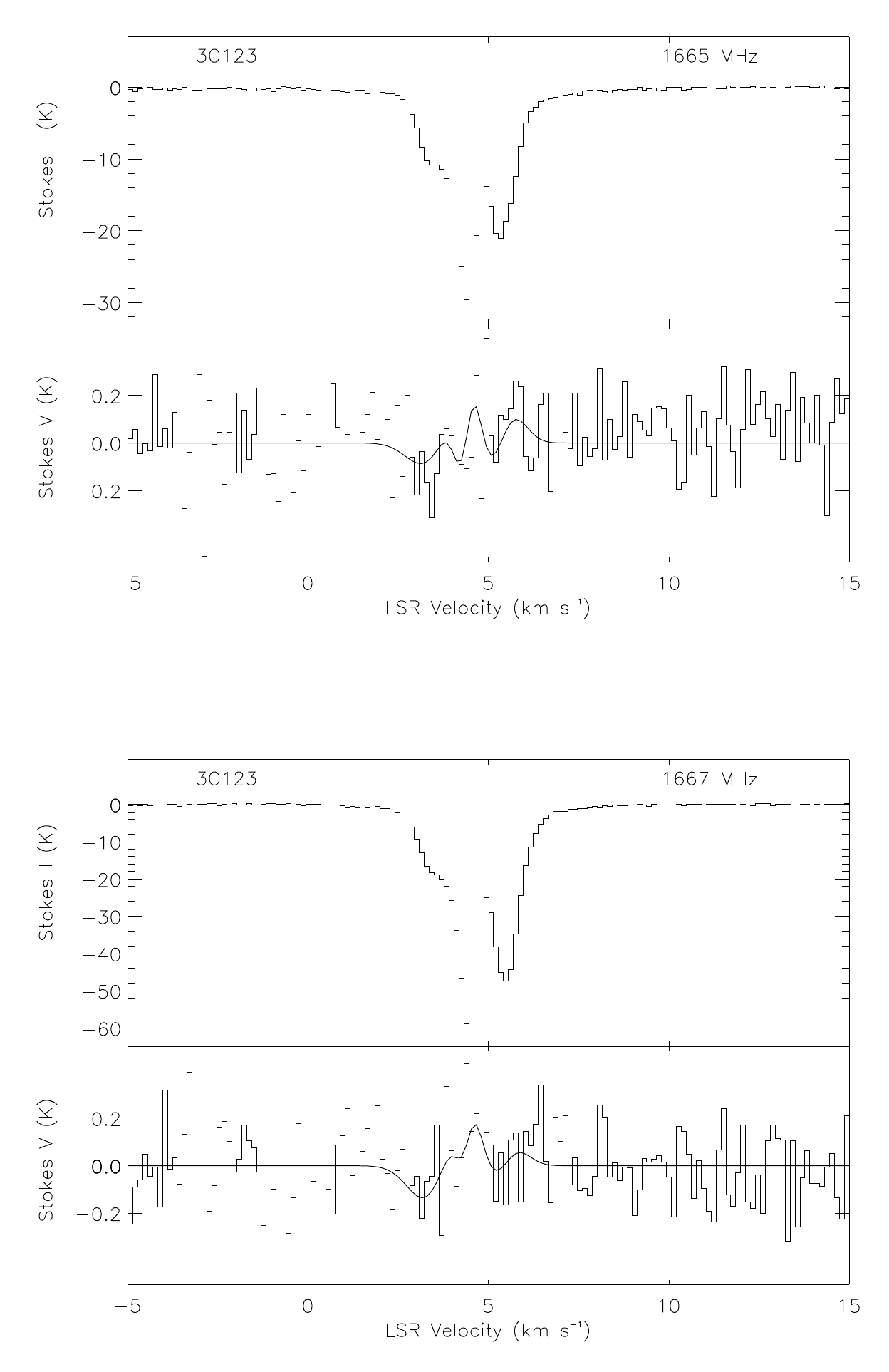}
\end{figure}

\begin{figure}[!h]
	\plotone{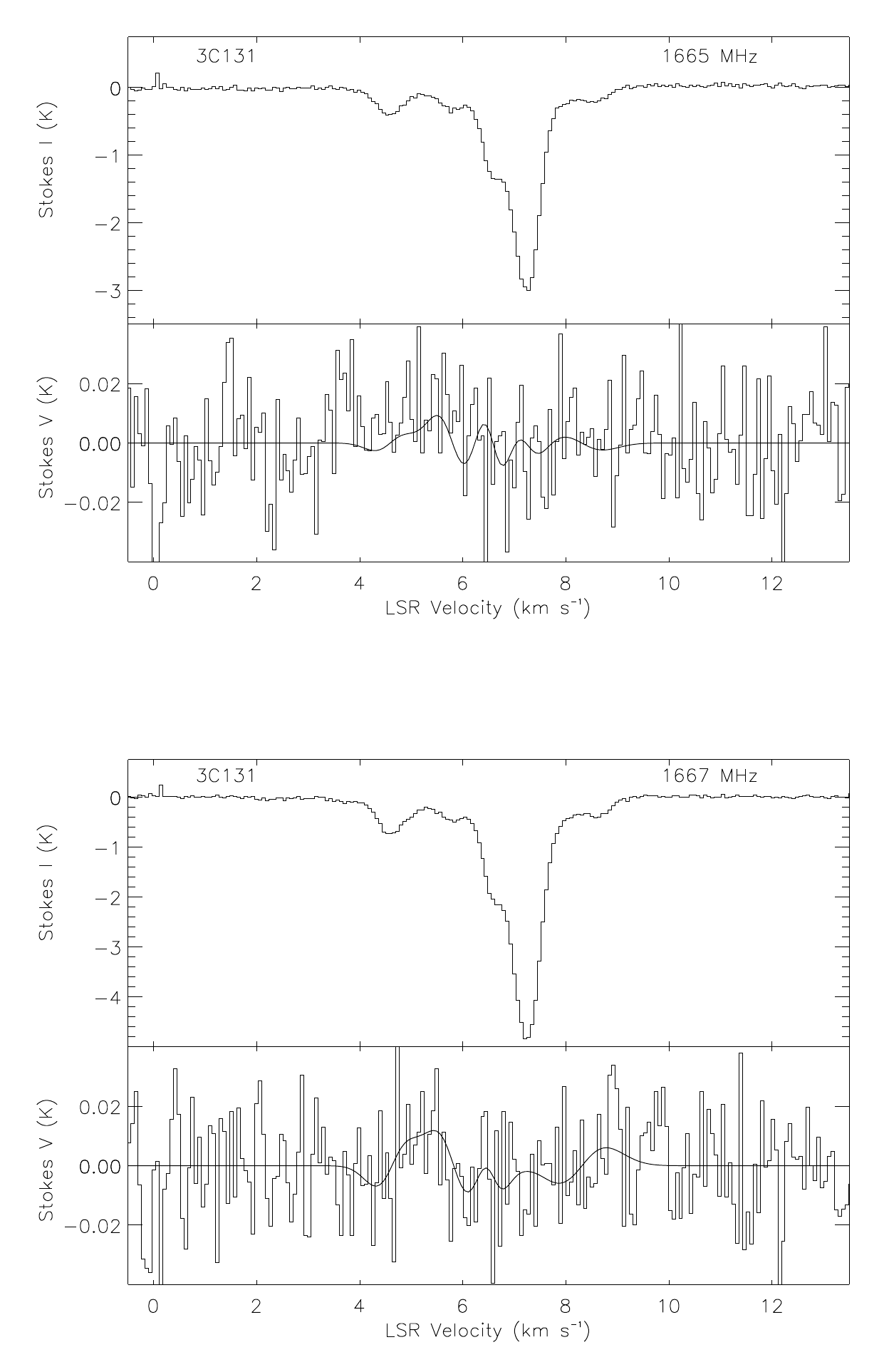}
\end{figure}

\begin{figure}[!h]
	\plotone{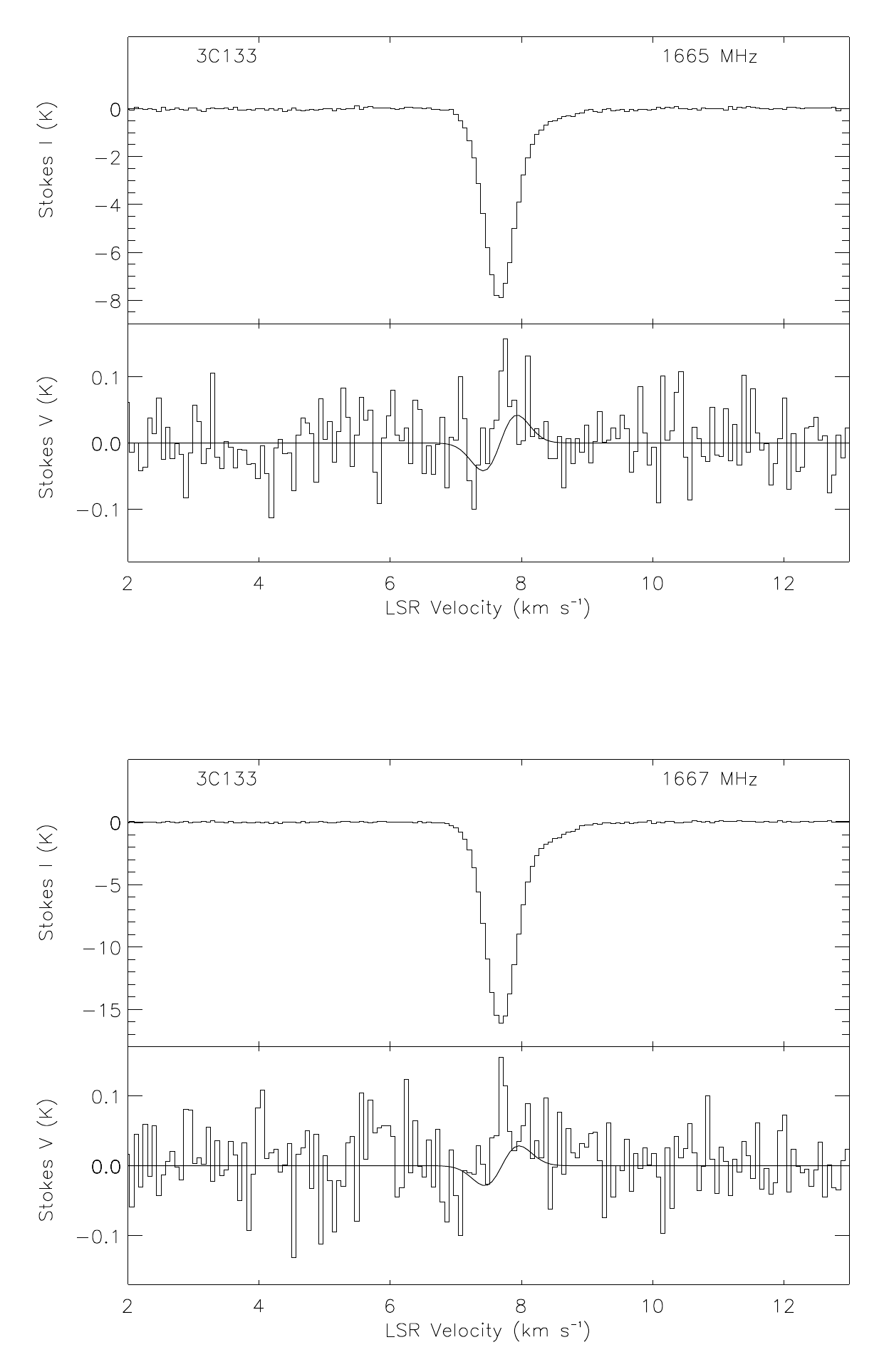}
\end{figure}

\begin{figure}[!h]
	\plotone{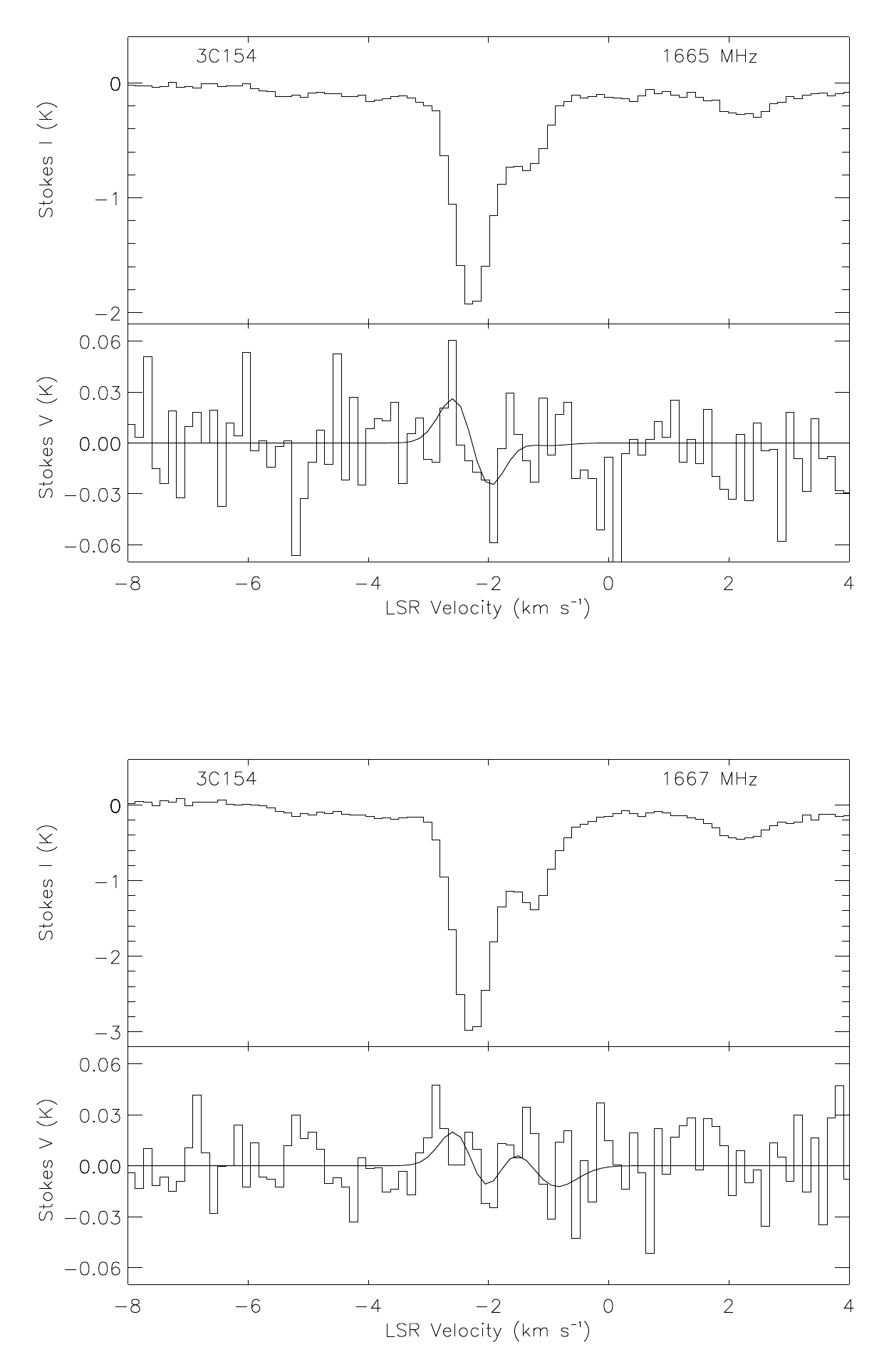}
\end{figure}

\begin{figure}[!h]
	\plotone{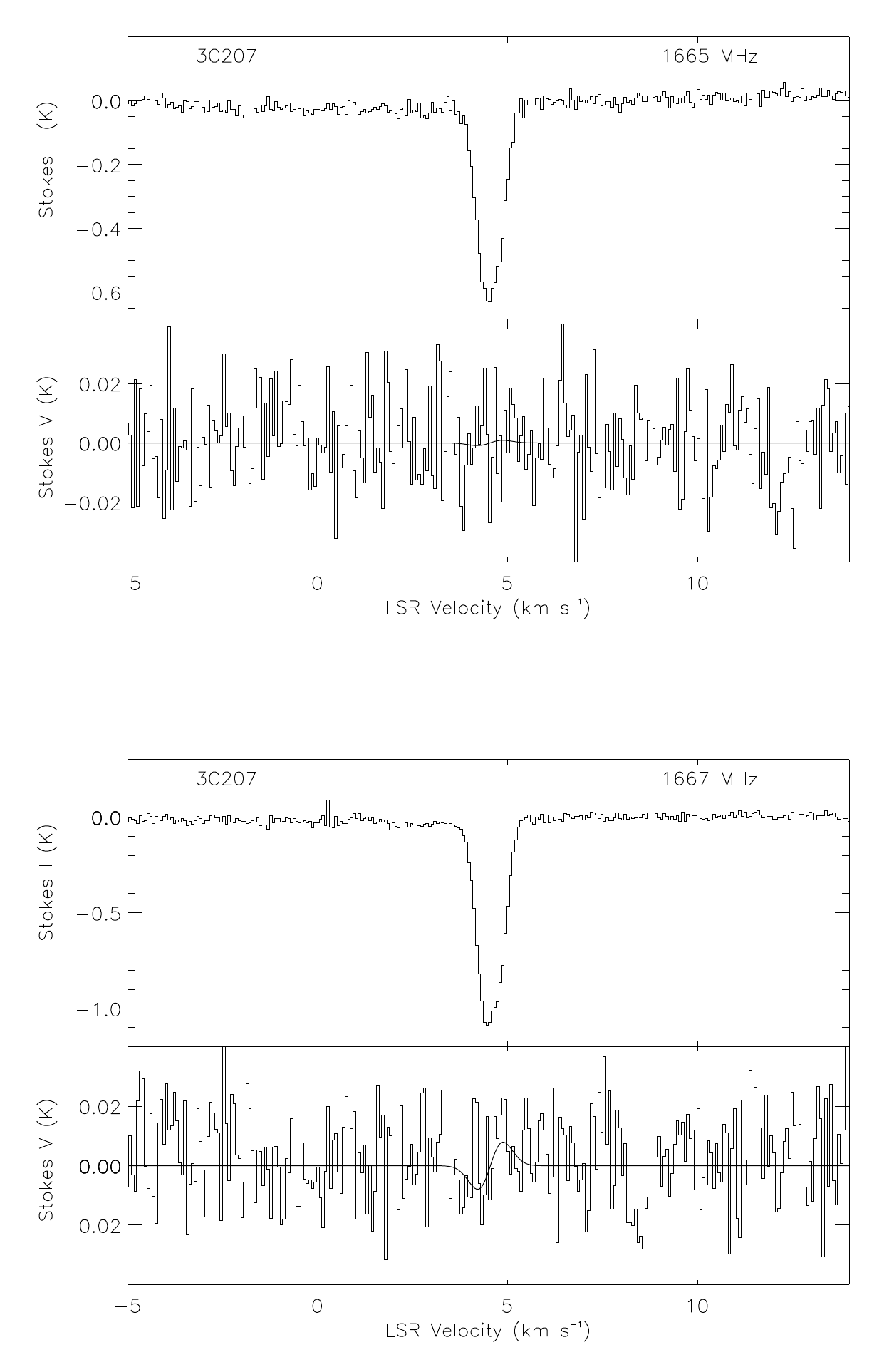}
\end{figure}

\begin{figure}[!h]
	\plotone{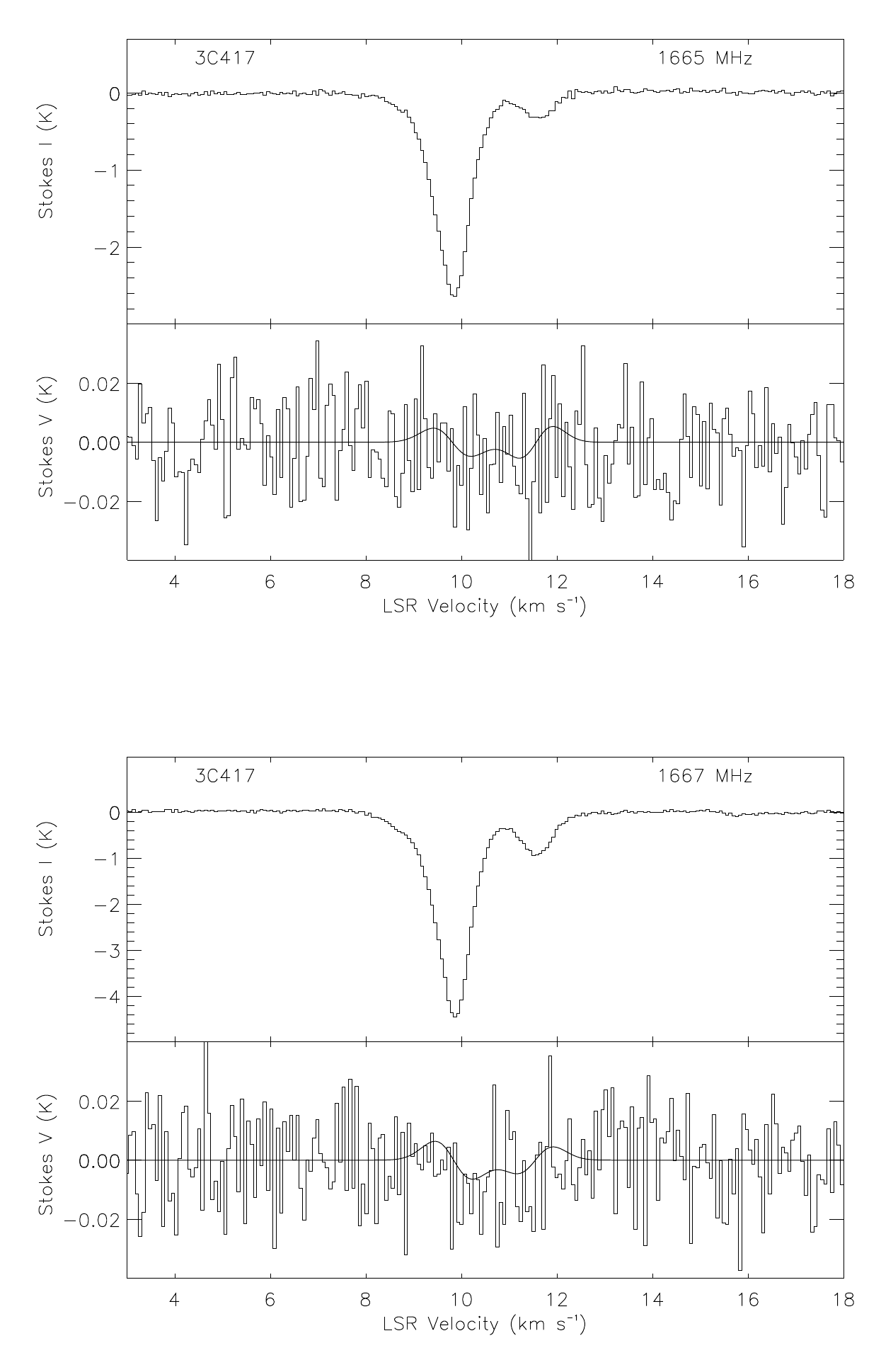}
\end{figure}

\begin{figure}[h]
	\plotone{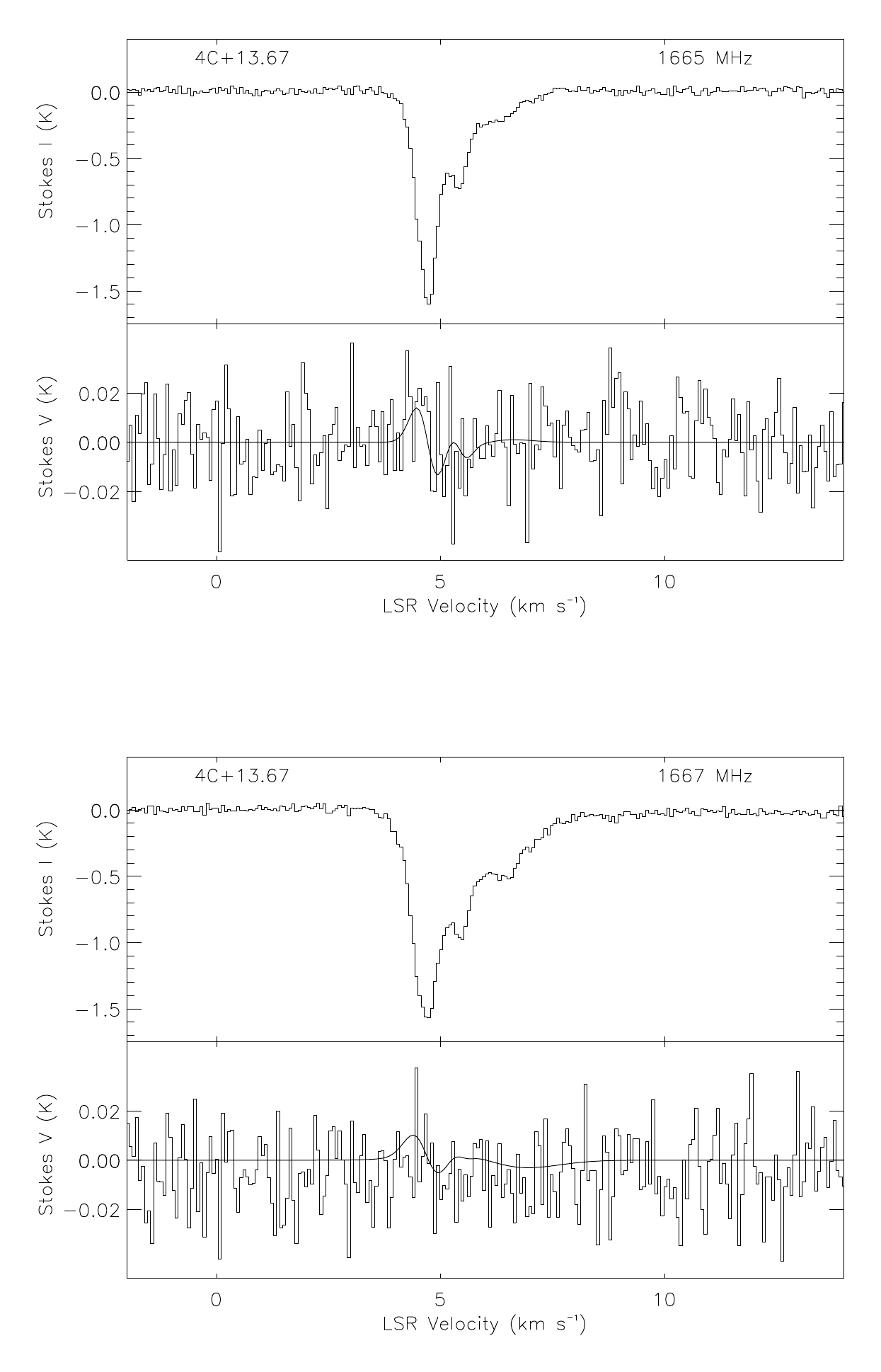}
\end{figure}

\begin{figure}[h]
	\plotone{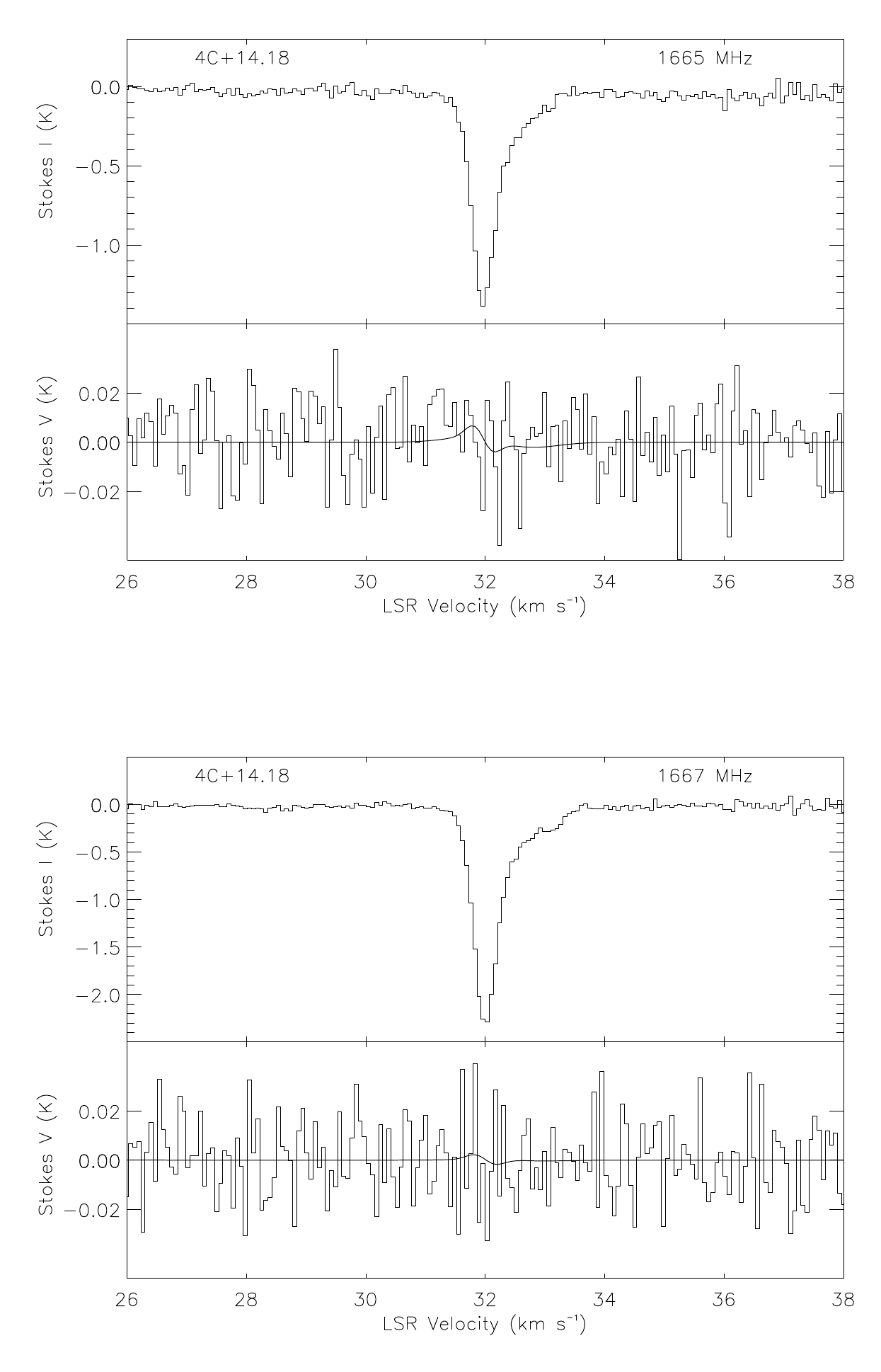}
\end{figure}

\begin{figure}[h]
	\plotone{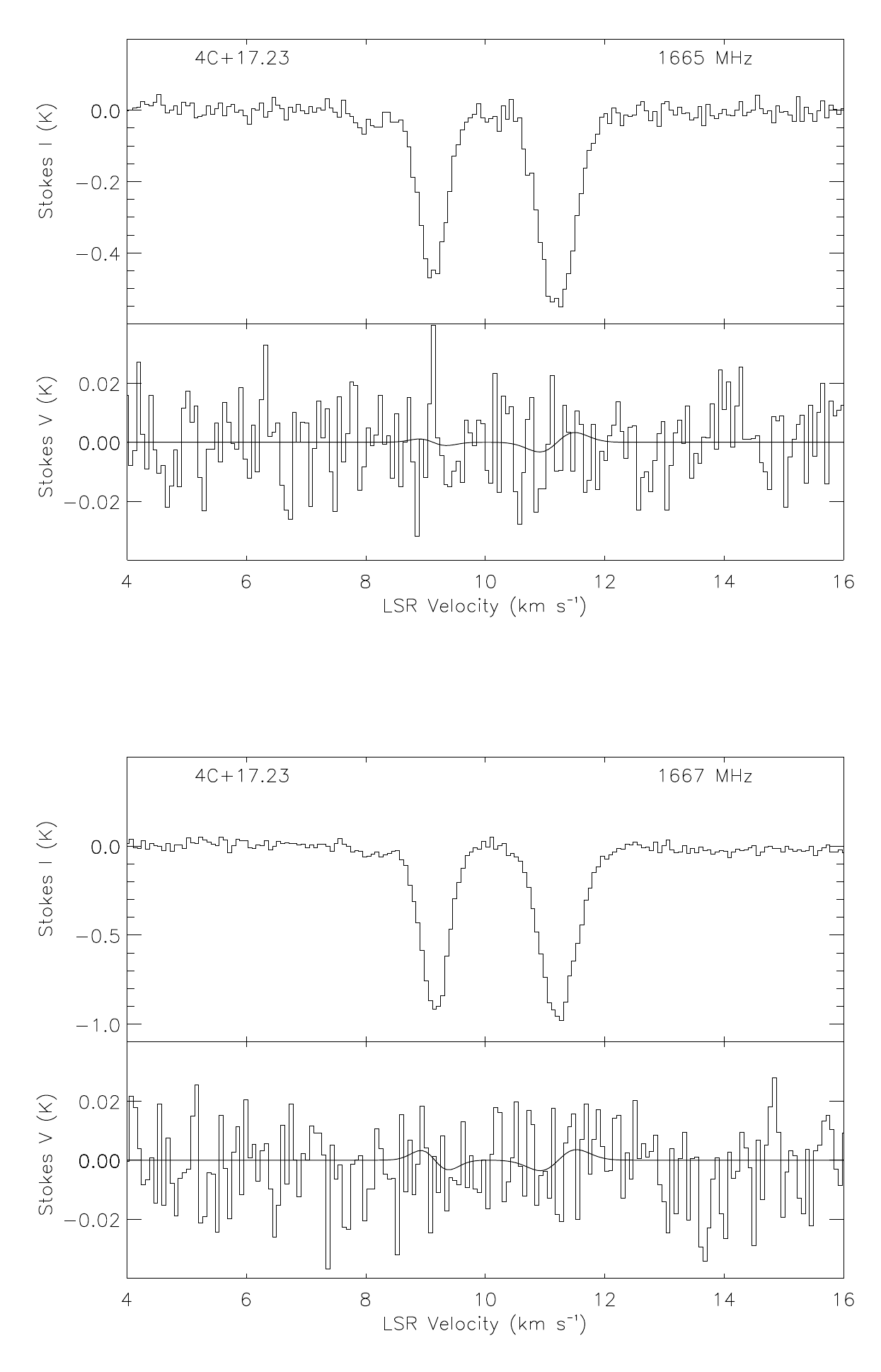}
\end{figure}

\begin{figure}[h]
	\plotone{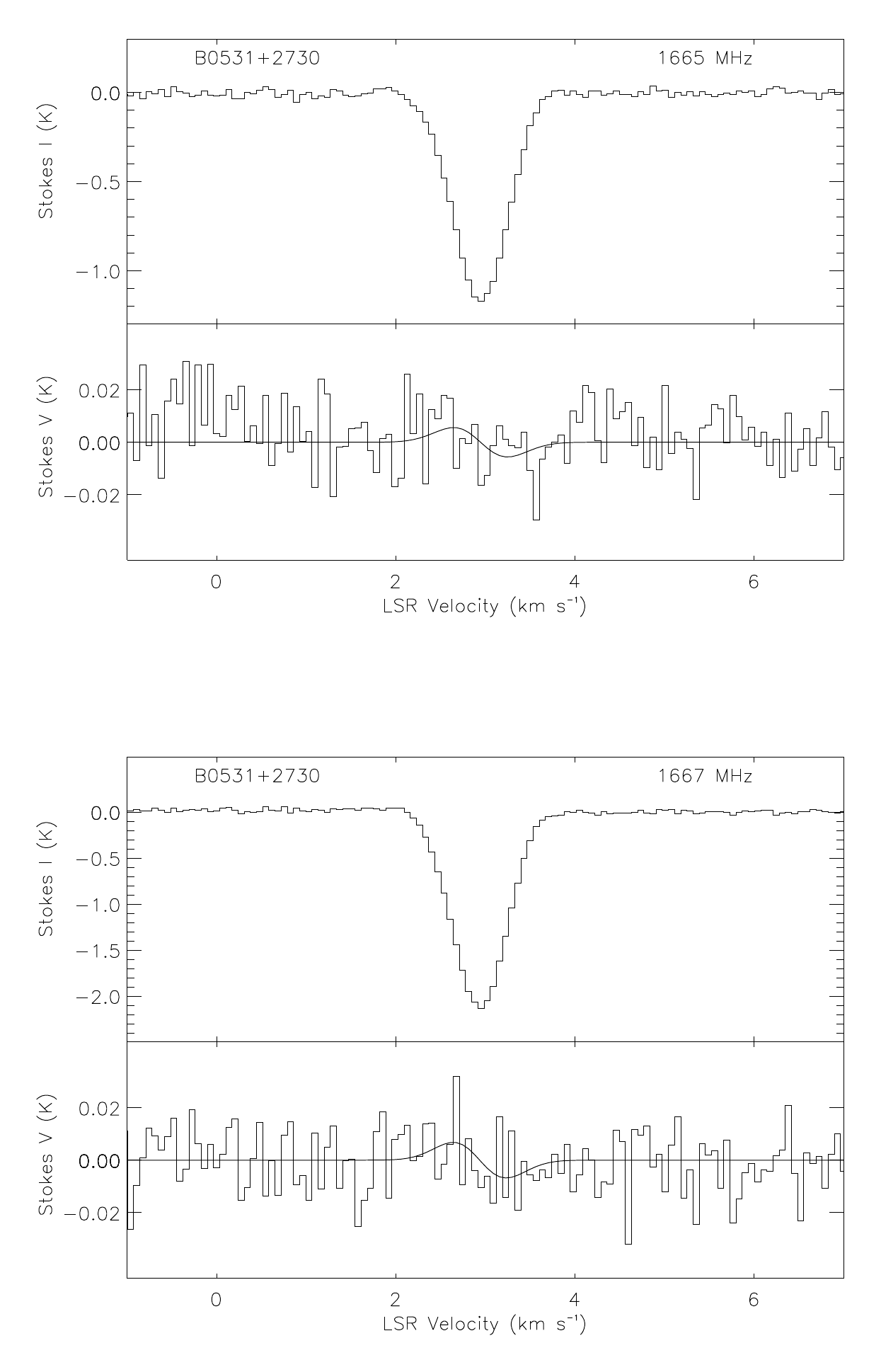}
\end{figure}

\begin{figure}[h]
	\plotone{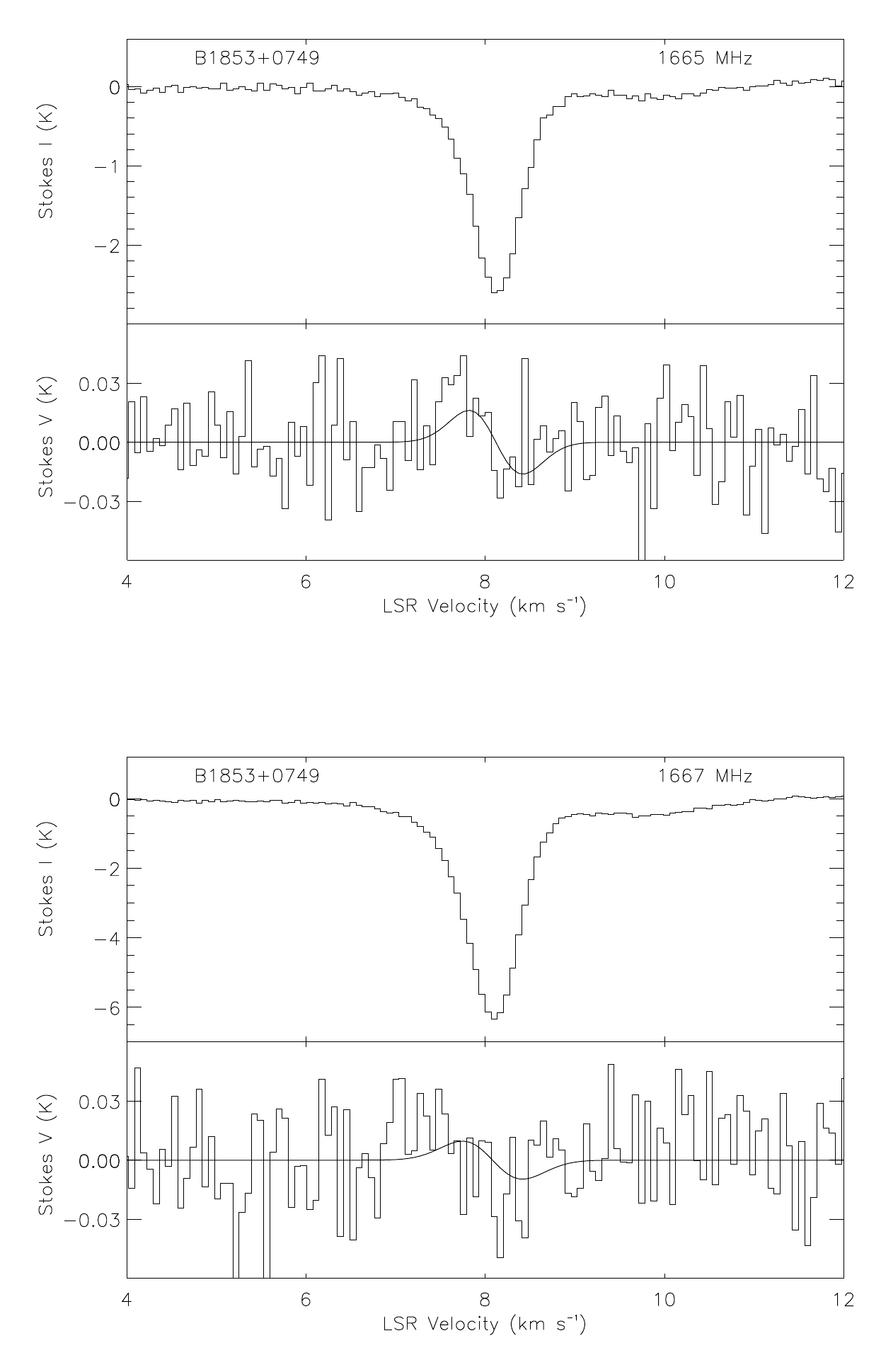}
\end{figure}

\begin{figure}[h]
	\plotone{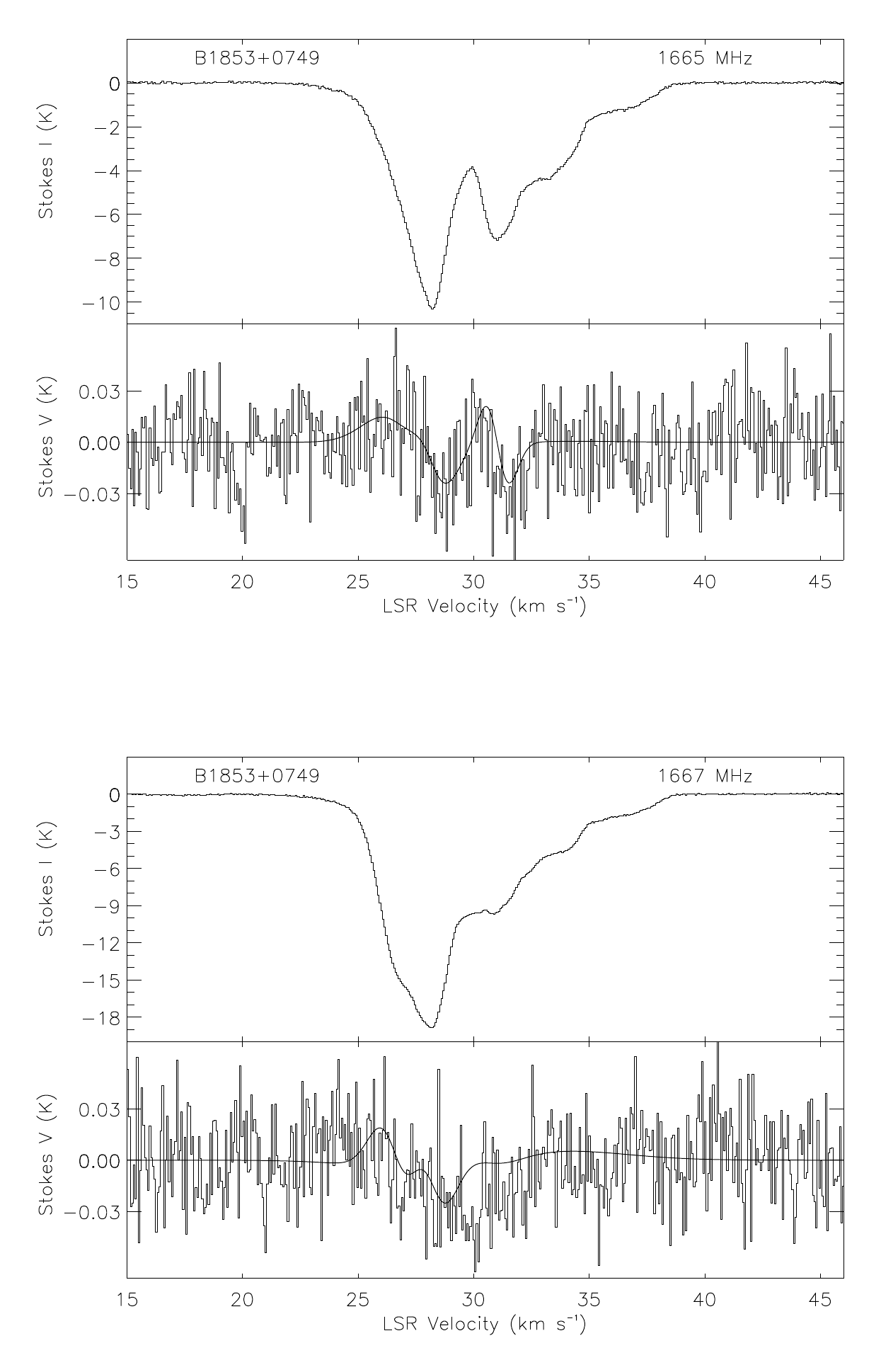}
\end{figure}

\begin{figure}[h]
	\plotone{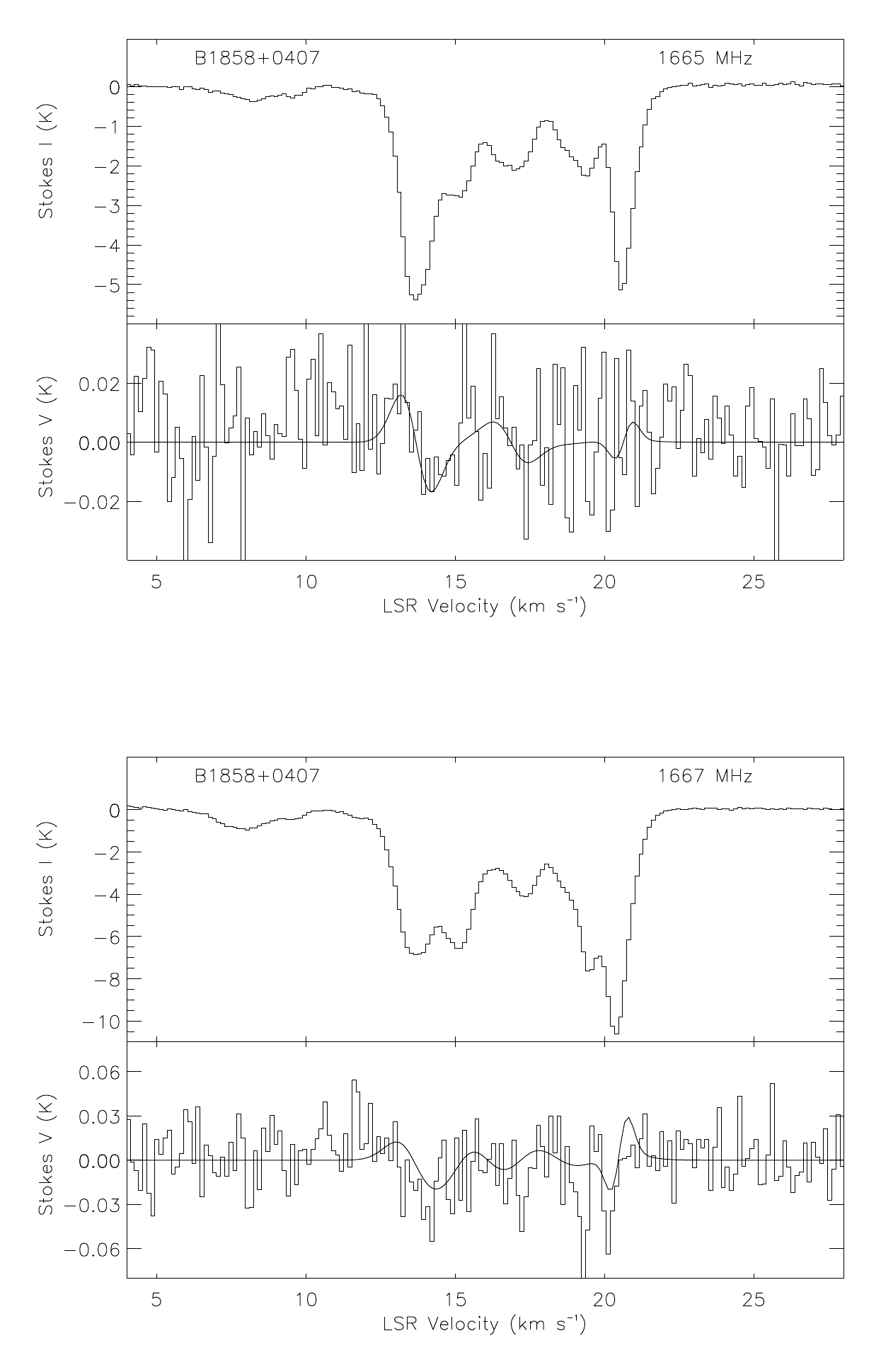}
\end{figure}

\begin{figure}[h]
	\plotone{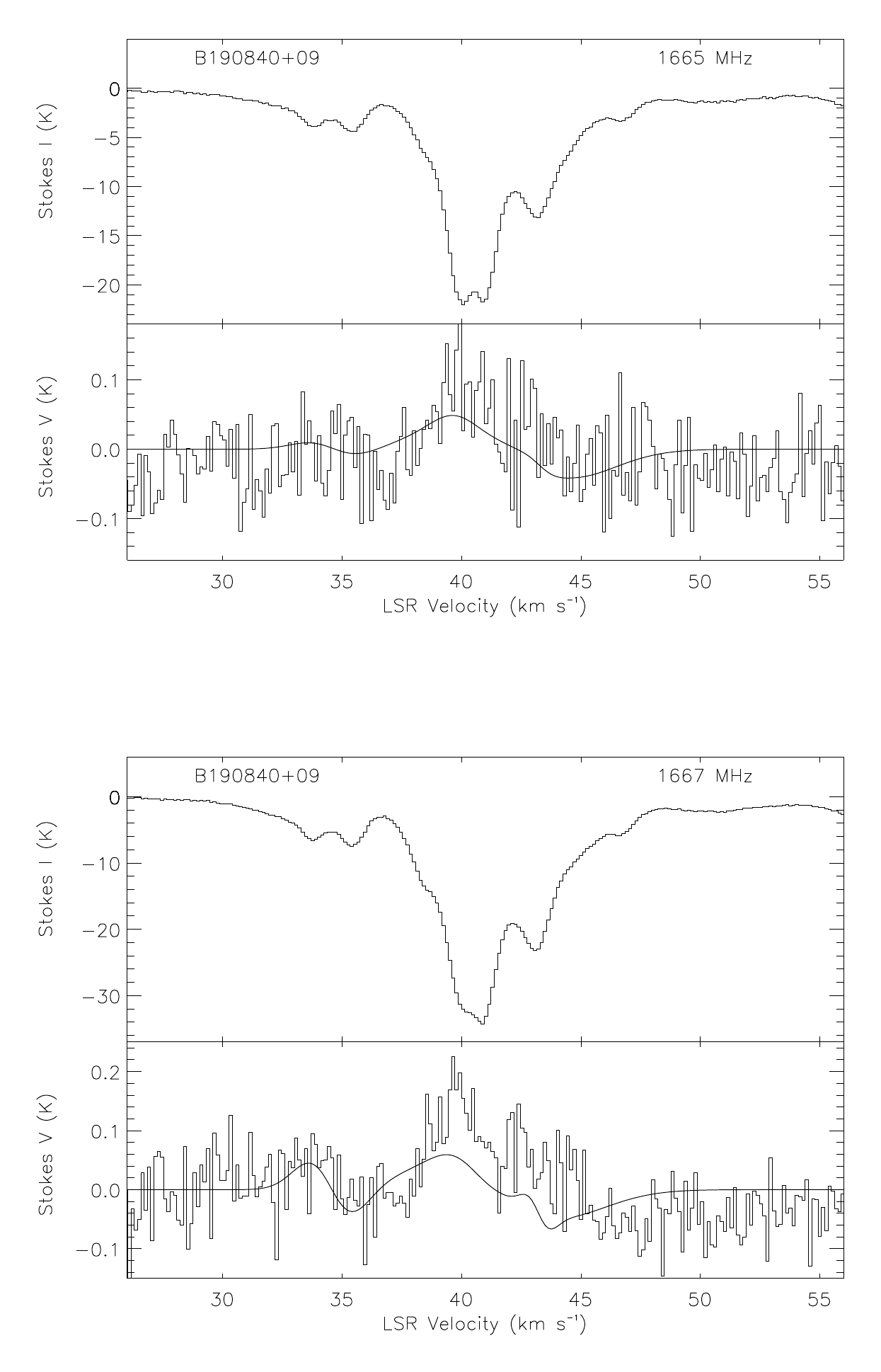}
\end{figure}

\begin{figure}[h]
	\plotone{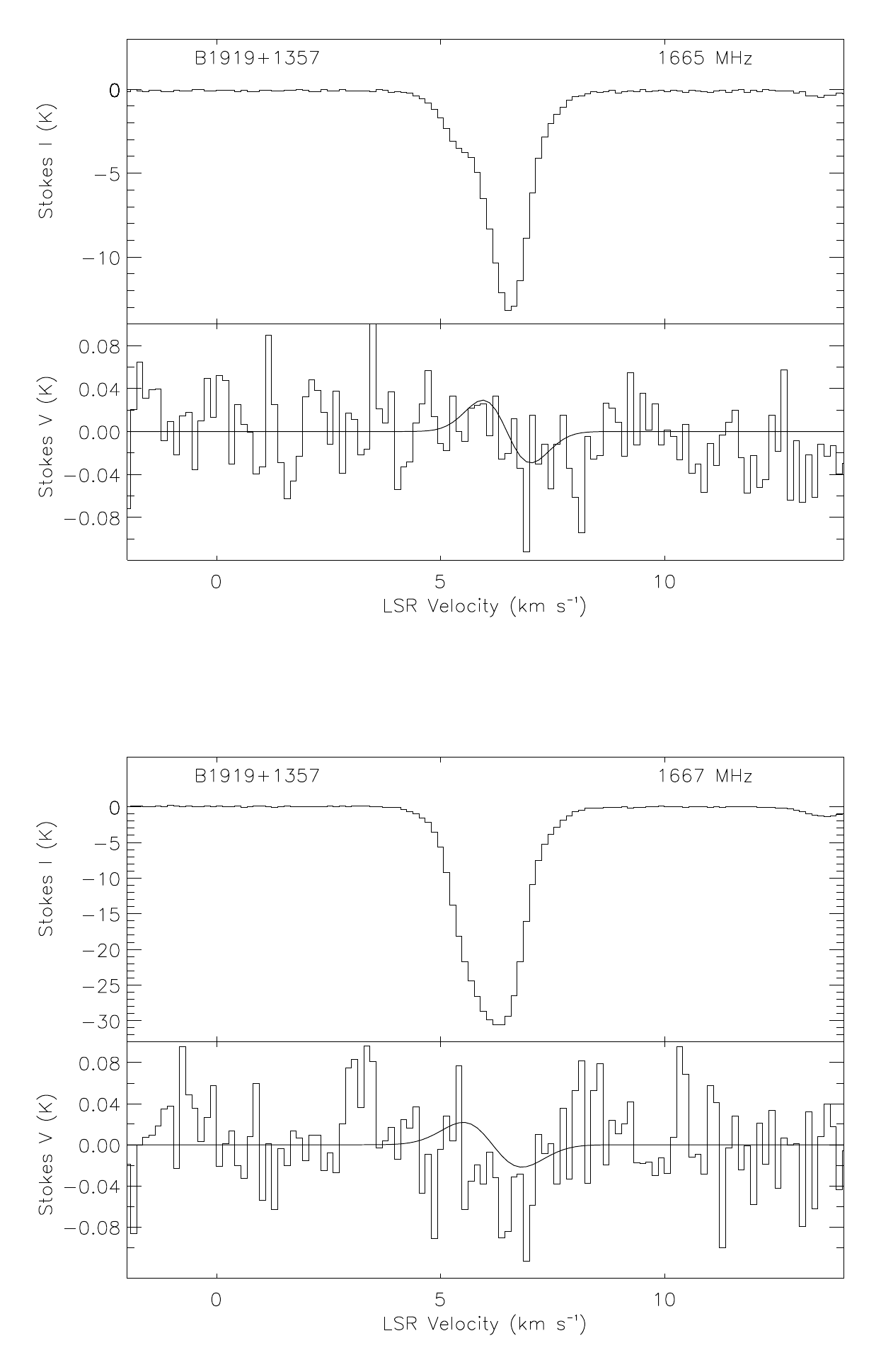}
\end{figure}

\begin{figure}[h]
	\plotone{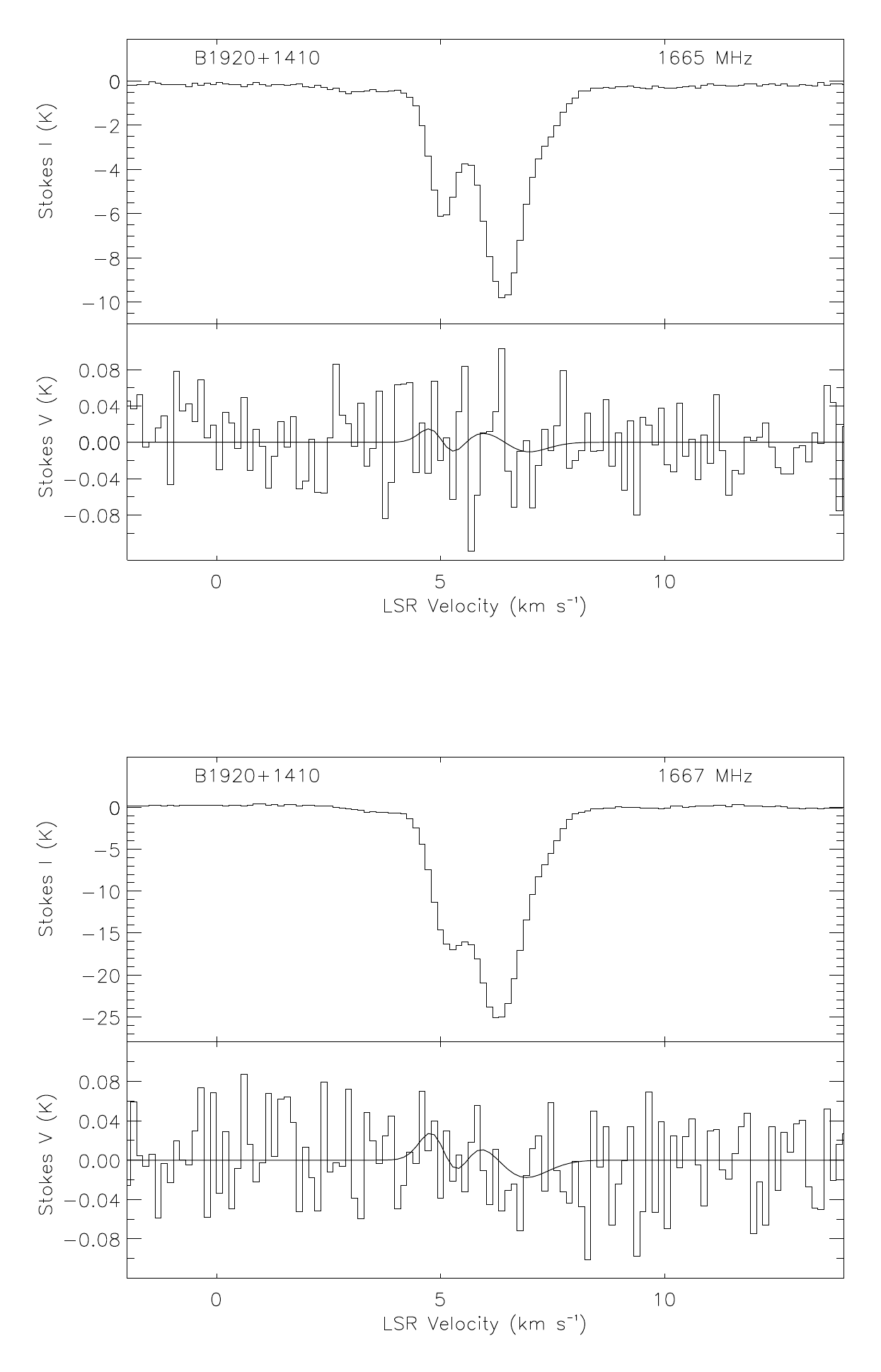}
\end{figure}

\begin{figure}[h]
	\plotone{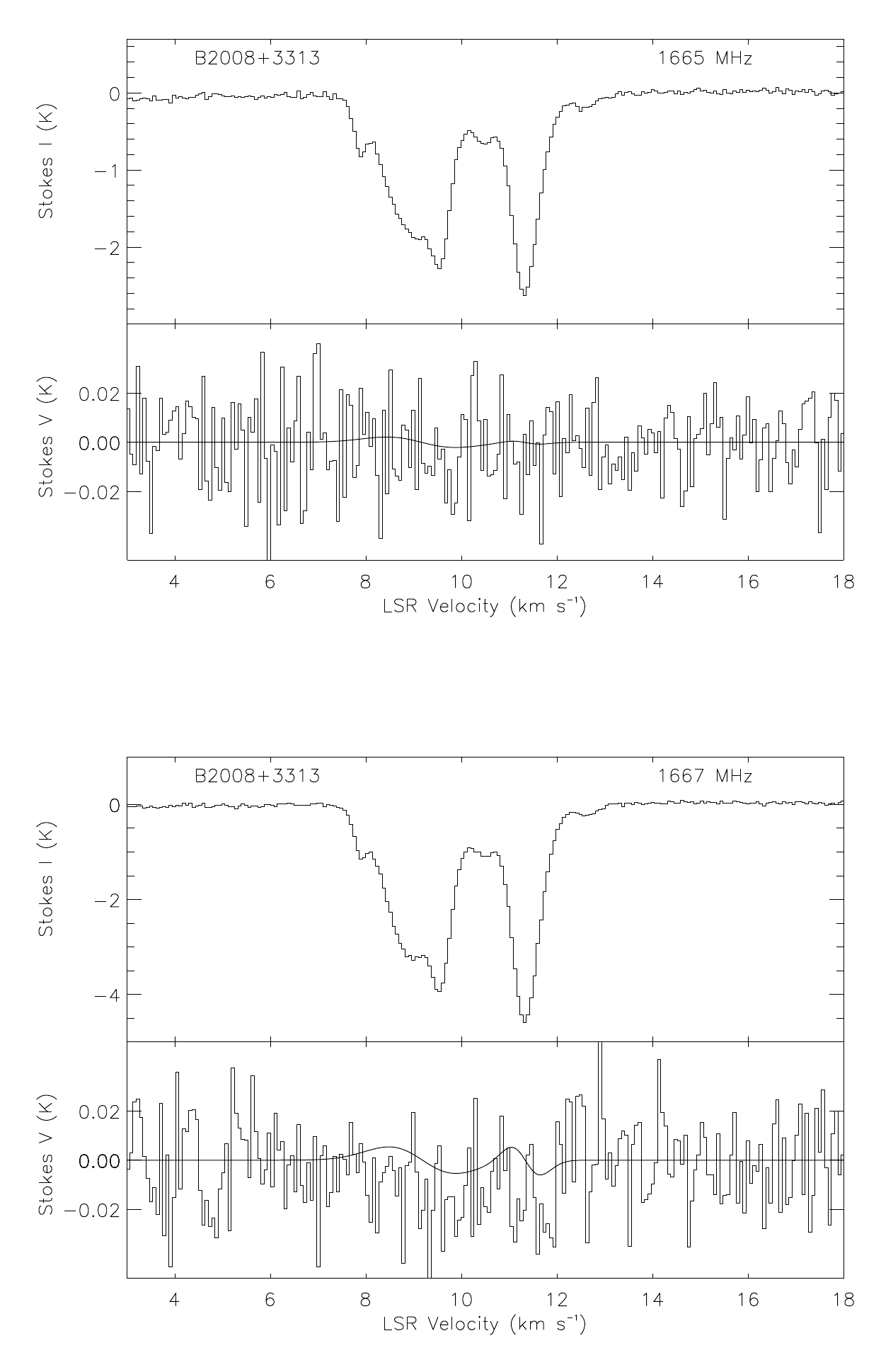}
\end{figure}

\begin{figure}[h]
	\plotone{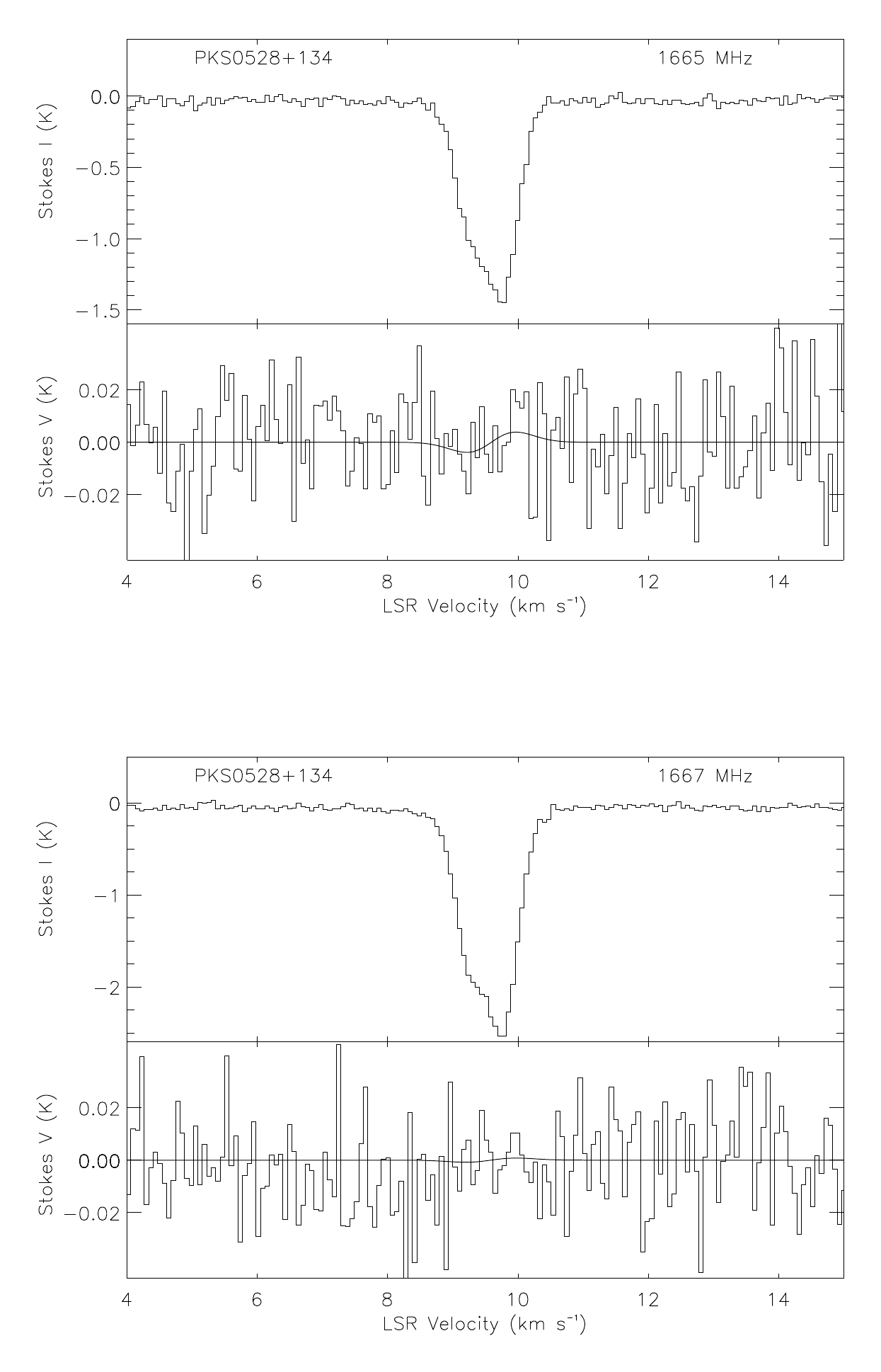}
\end{figure}

\begin{figure}[h]
	\plotone{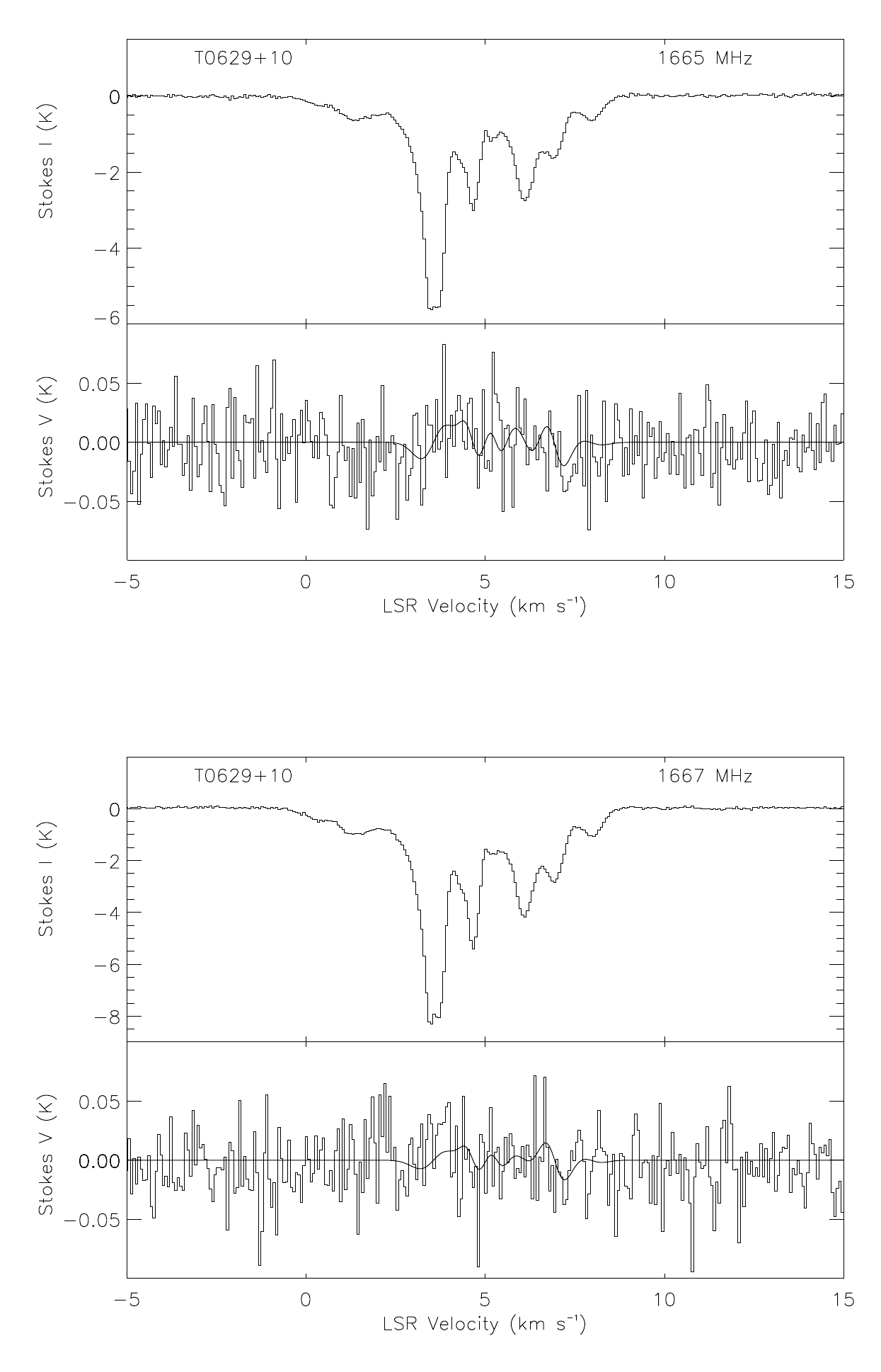}
\end{figure}

\clearpage

\end{document}